\documentclass[%
 aip,
 floatfix,
 amsmath,amssymb,
 reprint,%
]{revtex4-1}

\usepackage{graphicx}
\usepackage{amsfonts, amsmath, amstext, amssymb, amsfonts, amsxtra}
\usepackage{braket}
\usepackage{bbm}
\usepackage{bbold}
\usepackage{dcolumn}
\usepackage[protrusion=true,expansion=true]{microtype}
\usepackage{wasysym}
\usepackage{cleveref}
\usepackage{color}
\usepackage[normalem]{ulem}
\usepackage{bm}
\usepackage[utf8]{inputenc}
\usepackage[T1]{fontenc}
\usepackage{mathptmx}
\usepackage{float}
\usepackage{mathtools}

\newcommand{\id}{\mathbb{1}}  
\newcommand{\norm}[1]{\left\lVert#1\right\rVert}

\begin{document}

\preprint{AIP/123-QED}

\title{Quantum Lyapunov exponents and complex spacing ratios: \\ two measures of Dissipative Quantum Chaos}

\author{I.I.~Yusipov}
\email{yusipov.igor@gmail.com}
\affiliation
{Department of Applied Mathematics, Lobachevsky University, 603950, Nizhny Novgorod, Russia
}
\affiliation
{ 
    Mathematical Center, Lobachevsky University, 603950, Nizhny Novgorod, Russia
}

\author{M.V.~Ivanchenko}
\affiliation
{ 
    Department of Applied Mathematics, Lobachevsky University, 603950, Nizhny Novgorod, Russia
}

\date{\today}

\begin{abstract}
The agenda of Dissipative Quantum Chaos is to create a toolbox which would allow us to categorize open quantum systems into `chaotic' and `regular' ones. Two approaches to this categorization have been proposed recently. One of them is based on spectral properties of  generators of  open quantum evolution.  The other one utilizes  
the concept of Lyapunov exponents to analyze quantum trajectories obtained by unraveling this evolution.
By using two quantum models, we relate the two approaches and try to understand whether there is an agreement between the  corresponding categorizations. Our answer is affirmative.
\end{abstract}
\maketitle

\begin{quotation}
As its name hints, Dissipative Quantum Chaos (DQC) is a theory at the interface between classical Dissipative Chaos~\cite{Ott2002} and Quantum Chaos~\cite{Haake2018}. It it therefore natural that the emerging DQC toolbox includes methods adapted from these two well-developed theories.

One of the key concepts of  Quantum Chaos is universalities in the spectra of generators of unitary evolution, that are Hamiltonians. These universalities allow to classify Hamiltonians as chaotic or regular ones. Recently, the idea of spacing ratios\cite{Oganesyan2007,Pal2010,Atas2013,Buijsman2019}, an integral characteristic of Hamiltonian spectra, was generalized to the case of generators of quantum Markovian evolution~\cite{Sa2020}. Probability distributions of complex spacing ratios were suggested to be used as indicators of DQC and this suggestion was illustrated with a tunable many-body quantum model~\cite{Sa2020}.

The concept of Lyapunov exponents which allows to quantify instability of the  system dynamics by following the corresponding  trajectories in the system space space, is a pillar of classical Dissipative Chaos~\cite{Ott2002}. In a series of recent works, we generalized this concept to the dynamics of open quantum systems~\cite{Yusipov2019,Yusipov2020,Yusipov2021}. The idea is to unravel the evolution determined
by a Markovian generator into an ensemble of quantum trajectories and then to use
(in)stability characteristics of these trajectories as the basis to decide whether the original open quantum dynamics is regular (or chaotic). This idea was tested with several open models that allow for classical mean-field transition. We found correlations between chaos/order in the quantum case (quantified with Lyapunov exponents) and the type of the dynamics in the mean-field limit.

It is an open question whether the two approaches to DQC
are complementary and do not contradict each other. The aim of our work is to make a step towards getting the answer to this question.
\end{quotation}


\section{\label{sec:intro} Introduction}

There exist several approaches to model the evolution
of open quantum systems, i.e., systems that are interacting with their environment~\cite{Breuer2007}. One of the most popular is based on
the quantum master equation which governs evolution of the density operator $\rho$ of an open $N$-dimensional system, 
\begin{equation}
	\label{eq:master}
		\dot{\rho} = \mathcal{L}(\rho)
\end{equation}
with generator $\mathcal{L}$ of the so-called 
Gorini-Kossakowski-Sudarshan~\cite{Gorini1976} and Lindblad~\cite{Lindblad1976} form (henceforth \emph{Lindbladian})~\cite{remark},
\begin{equation}\label{L}
  \mathcal{L}(\rho) = -i[H,\rho] + \sum_k \gamma_k \Big(L_k \rho L_k^\dagger - \frac 12 \{ L_k^\dagger L_k,\rho\} \Big),
\end{equation}
where $\{... \}$ denotes anti-commutator, $H^\dagger = H$ is the system Hamiltonian, $L_k$ are jump operators (they capture incoherent action of the environment on the system), and all rates are positive, $\gamma_k > 0$. Being given all these operators and rates, one should be able to answer the following question: Is the evolution induced by $\mathcal{L}$  chaotic?

The question is similar to the one in Quantum Chaos forty years ago, when, being given Hamiltonian $H$, one had to determine whether the evolution induced by this operator, $\vert \dot \psi \rangle = iH\vert \psi \rangle $, is regular or chaotic. It turned out that the answer is encoded in the spectrum of $H$ and it can be retrieved  by using Random Matrix Theory (RMT)~\cite{Mehta2004}. 

More specifically, one has to investigate the behavior of correlations between eigenvalues (energies) of the Hamiltonian. These correlations are described with the  probability density function $P(s)$ of spacing between consecutive energies, $s_j = \varepsilon_{j+1} - \varepsilon_{j}$~\cite{Mehta2004}. One of the landmark results of RMT is the power-law level repulsion $P(s) \propto s^{-\beta}$ in the limit $s \rightarrow 0$, with exponent values specific to the three main Gaussian ensembles that are referred to as Gaussian Orthogonal Ensemble 
(GOE, $\beta = 1$), Gaussian Unitary Ensemble (GUE, $\beta = 2$), and Gaussian Symplectic Ensemble (GSE, $\beta = 4$)~\cite{Mehta2004}. The first one, GOE, is the most relevant in the context of Quantum Chaos. Whenever the Hamiltonian of a system yields $P(s)$ close to the  distribution typical of the matrices from this ensemble (the so-called Wigner-Dyson distribution), the system is identified as chaotic~\cite{Haake2018}. The spectra of `regular' Hamiltonians, in contrast, are characterized by exponential (Poisson) distributions.
The notion of generalized Gaussian $\beta$ ensemble~\cite{Dyson1962}, with $0 \leq \beta \leq 1$, allows to quantify the transition between chaotic and regular regimes in a continuous manner~\cite{Guhr1998}.

In practice, when analyzing spectra of model Hamiltonians, one has to eliminate the dependence on fluctuating local energy density and only then compare the obtained distributions with Poisson  or  Wigner-Dyson distributions. For that a complicated  unfolding procedure \cite{Haake2018} has to be performed. It can be avoided by using \textit{ratios} of consecutive spacing \cite{Oganesyan2007}. Expressions for spacing ratio (SR) distributions for different RM ensembles were derived \cite{Atas2013} and currently these distributions are popular tools to analyze many-body Hamiltonians~\cite{Pal2010,Atas2013,Buijsman2019}.

The notion of spacing ratios was generalized to the case of Lindbladians in a recent work by S{\'a}, Ribeiro, and Prosen ~\cite{Sa2020}.
The eigenvalues  $\lambda_k$ of Lindbladian 
are complex numbers and consecutive eigenvalues have to be ordered based on the distance on the complex plane. Then the complex spacing ratios (CSR) can be calculated and their probability distribution over the complex plane can be sampled. By using an open  many-body model, it was demonstrated that, by varying model parameters, it is possible to tune
CSR distributions from the distributions exhibited by members of the Gaussian Ginibre Unitary Ensemble (GinUE)~\cite{Ginibre1965} to the ones  characteristic to diagonal matrices with complex Poisson-distributed entries. 
Lindbladians exhibiting  CSR distributions typical of GinUE ensemble were coined `chaotic', while the ones with the distributions typical of diagonal random matrices were coined `regular'.

In our recent works~\cite{Yusipov2019,Yusipov2020,Yusipov2021}, we proposed an alternative way to
quantify the degree of chaos in the evolution determined by Lindbladian $\mathcal{L}$. Namely, we introduced a quantum  version of
Lyapunov exponents (LEs) which is based on the idea of unraveling the master equation, Eqs.~(1-2), into an ensemble of stochastic processes called `quantum trajectories'~\cite{Plenio1998,Daley2014}. The evolution of every quantum trajectory is governed by a non-Hermitian operator (`effective Hamiltonian') in a continuous-time manner and interrupted by random jump-like events (that are actions of the jump operators). We illustrated this idea with several models that permit semiclassical description and demonstrated that there is an agreement between the type of the mean-filed dynamics and the sign of the largest LE obtained for the quantum model.  

In this paper we consider two genuinely quantum models which do not allow for semiclassical transition and have no classical mean-field versions. We analyze their spectra and calculate the corresponding quantum Lyapunov exponents. We demonstrate that, at least for these models, there is an agreement between the two measures of DQC.

\section{\label{sec:methods} Methods}

In this section we provide some basic information on the two approaches to characterize Dissipative Quantum Chaos.

\subsection{\label{sec:methods:ratio} Complex spacing ratios}

To calculate this measure for a model Lindbladian, we first diagonalize $\mathcal{L}$ and find its eigenvalues, $\left\{ \lambda_k \right\}_{k=1}^N$. Next,  for each $\lambda_k$ we find its nearest neighbor ($NN$) eigenvalue, $\lambda_k^{NN}$, and next to-nearest neighbor ($NNN$) eigenvalue $\lambda_k^{NNN}$ by using the standard Euclidean norm. For this triple we then calculate the complex spacing ratio (CSR)
\begin{equation}
	\label{eq:csr}
	\begin{gathered}
	    z_k = \frac{\lambda^{NN}_k - \lambda_k}{\lambda^{NNN}_k - \lambda_k}.
	\end{gathered}
\end{equation}
CSR values are confined to the unit disc and the corresponding probability density function (pdf)  $P_{\mathrm{CSR}}(z) \coloneqq  P_{\mathrm{CSR}}[\mathrm{Re}(z),\mathrm{Im}(z)]$, 
has this disc as a support. In Ref.~\cite{Sa2020} this distribution was used to categorize  many-body Lindbladians as `chaotic' and `regular' ones. 
It was observed that non-integrable  Lindbladians yield CSR distributions similar to the one exhibited by members of Gaussian Ginibre Unitary Ensemble (GinUE)~\cite{Ginibre1965}, while the integrable ones produce CSR distributions characteristic of diagonal matrices with complex Poisson-distributed entries. In the former case the eigenvalues are correlated and this results in a distinctive  horse-shoe pattern with depletion regions near $z=0$ and $z=1$, whereas  in the latter case the eigenvalues are independent and, in the  limit $N \rightarrow, \infty$  $P_{\mathrm{CSR}}(z)$ tends to the flat distribution over the unit disc.

There are several ways to compare a histogram of the CSRs sampled with a finite-dimensional Lindbladian and the two characteristic probability density functions. In the original work~\cite{Sa2020}, this comparison was mainly visual since specific quantification was not the issue (the emphasis was made on elaboration of the general theoretical framework). 
Nevertheless,  some possibilities to quantify the distance between the distributions were suggested, e.g., by using marginalized one-dimensional distributions (cf. Section IV).  

\subsection{\label{sec:methods:qle} Quantum Lyapunov Exponent}

Here we outline the idea of quantum LEs (cf. Refs.~\cite{Yusipov2019,Yusipov2020,Yusipov2021} for a more detailed description).

As it was already mentioned in the introduction, the calculation is based on the procedure of unraveling \cite{Plenio1998,Daley2014} (also known as "Monte Carlo wave-function (MCwf) method" \cite{Dum1992, Molmer1993}), which replaces the evolution of the density operator governed by the Markovian master equation (1-2) with an ensemble of quantum trajectories. 

The evolution of the model system is now described by an ensemble of pure states, $\lbrace ...,\left|\psi(t)\right\rangle,...\rbrace$, with the evolution of every member of this ensemble governed by an effective non-Hermitian Hamiltonian, 
\begin{equation}
	\label{eq:qj_hamiltonian}
	\begin{gathered}
	\left|\dot{\psi}(t) \right\rangle  = iH_{eff} \left|\psi(t) \right\rangle =i (H -\frac{i}{2} \sum_{k=1}^{M-1}\gamma_k kL_k^\dagger L_k) \left|\psi(t) \right\rangle.
	\end{gathered}
\end{equation}
Because of non-Hermiticity of the Hamiltonian, the norm $||\psi(t)||$ of the wave function is monotonously decreasing in time and, as it reaches a threshold $\eta$, renewed each time from a set of independent random numbers uniformly distributed on the unit interval $[0,1]$, a jump is performed by acting on the wave function with one of the jump operators $L_k$'s (the choice of a particular operator is random and specified by a set of $M$ probabilities; see Ref.~\cite{Breuer2007} for more detailed description of the procedure). After that the norm is reset to one, $||\psi(t)||=1$, (by scaling the wave function) and the continuous non-unitary evolution described by  Eq.~(\ref{eq:qj_hamiltonian}), is initiated  again, until the next quantum jump occurs, etc.

The largest quantum Lyapunov exponent \cite{Yusipov2019} is calculated as the average rate of exponential growth of the distance between the base $\psi_{b}(t)$ and perturbed $\psi_v(t)$ quantum trajectories that evolve according to the Eq.~(\ref{eq:qj_hamiltonian}), in full analogy with the classical definition \cite{Benettin1976}.

The distance between the trajectories is determined by using the expectation value $o(t)$ of some observable (Hermitian operator) $O$, 
\begin{equation}
	\label{eq:le_obs}
	\begin{gathered}
		o(t) =  \langle \psi(t)| O | \psi(t) \rangle,
	\end{gathered}
\end{equation}
As it has been shown in our previous works, the values of the average Lyapunov exponent do not dependent much on a particular choice of the observable (safe for degenerate cases) \cite{Yusipov2019, Yusipov2020}. There, f.e., we used  energy of the system as such an observable so that $O = H$.

The perturbed trajectory $\psi_v(t)$ is initialized as a normalized perturbation vector: 
\begin{equation}
	\label{eq:le_var_tr}
	\begin{gathered}
		\left|\psi_v(t)\right\rangle = \frac{\left|\psi_b(t)\right\rangle + \varepsilon  \frac{\left|\psi_r\right\rangle}{\norm{\left|\psi_r\right\rangle}}}{\norm{\left|\psi_b(t)\right\rangle + \varepsilon \frac{\left|\psi_r\right\rangle}{\norm{\left|\psi_r\right\rangle}}}} \norm{\left|\psi_v(t)\right\rangle},
	\end{gathered}
\end{equation}
where $\left|\psi_r\right\rangle$ is a random wave function whose amplitudes are uniformly distributed on the interval $\left[-1, 1\right]$ and $\varepsilon$ is the magnitude of the difference between the trajectories.

The value of $\varepsilon$ is determined in such a way that the difference between two expectation values of the observable, calculated for the base, $\psi_{b}(t)$, and perturbed, $\psi_v(t)$, trajectories, is equal to a fixed value $\Delta(t) = \Delta_0$: 
\begin{equation}
	\label{eq:le_var_obs}
	\begin{gathered}
		\Delta(t) = \left|o_b(t) - o_v(t) \right|.
	\end{gathered}
\end{equation}
The difference between the current deviation and its target value $\eta = \Delta - \Delta_0$ is a monotonic function of the perturbation $\varepsilon$. 
The bisection algorithm is used to find an optimal value of $\varepsilon$, which minimizes the value of $\left| \eta \right|$.

During the time evolution of both trajectories with Eq.~(\ref{eq:qj_hamiltonian}), the renormalization of the perturbed trajectory is performed at equidistant instants of time, $t_k = k \tau$, $k \in \mathbb{N}$. At these instances, the local growth factor is calculated:
\begin{equation}
	\label{eq:le_growth_factor}
	\begin{gathered}
		d_k = \frac{\Delta(t_k)}{\Delta_0},
	\end{gathered}
\end{equation}
which is then used to calculate the largest Lyapunov exponent:
\begin{equation}
	\label{eq:le}
	\begin{gathered}
		\lambda=\lim\limits_{t \to \infty} \frac{1}{t} \sum\limits_k \ln d_k.
	\end{gathered}
\end{equation}
A particular choice of  $\tau$ does not  affect the values of the quantum Lyapunov exponent significantly, provided that it is chosen within some reasonable window~\cite{Yusipov2019}.

\section{\label{sec:model} Models}

\subsection{\label{sec:model:integrable} Integrable Lindbladian}

Recently, a method to construct  integrable Lindbladians has been proposed~\cite{deLeeuw2021}. 
The method is based on `quantization' of exactly solvable classical continuous-time Markov processes. Out of several examples presented in Ref.~~\cite{deLeeuw2021} we  use the model listed there as `model B1'. It describes a one-dimensional spin-1/2 chain consisting of $M$ spins. Its Lindbladian has no unitary part, $H=0$, and each one of its $M-1$ jump operators, $L_l$, acts in the subspace of two neighboring spins, $l$ and $l+1$, $l=1,...,M$,
\begin{equation}
 L_l =
 \label{eq:integ}
  \begin{pmatrix}
    \eta & 0 & 0 & 0  \\
    0 & 0 & 1 & 0 \\
    0 & 1 & 0 & 0 \\
    0 & 0 & 0 & \kappa 
  \end{pmatrix},
\end{equation}
where $\eta = \pm 1$ and $\kappa = \pm 1$. We set $\eta = 1$ and $\kappa = -1$. 

\subsection{\label{sec:model:MBL} Open many-body system}

As the second model we use an open version of a many-body system exhibiting the phenomenon of many-body localization (MBL)~\cite{Basko2006}.

The model system  represents a chain of $M$ (an even number) sites, occupied by $M/2$ spinless fermions. The systems evolves in the half-filling subspace with the total number of states $S=M!/\left[ \left(M/2\right)!\right]^2$.
The model Hamiltonian has the form~\cite{Pal2010}
\begin{equation}
	\label{eq:Hamiltonian}
	\begin{gathered}
		H = W \sum_{l=1}^{M} h_l n_{l} - J \sum_{l=1}^{M-1}\left(c^{\dagger}_l c_{l+1} + c^{\dagger}_{l+1} c_{l}\right) + \\ + U \sum_{l=1}^{M-1} n_{l} n_{l+1} , 
	\end{gathered}
\end{equation}
where $c^\dagger_l$ ($c_l$) creates (annihilates) a fermion at site $l$, and $n_l=c^\dagger_l c_l$ is the local particle number operator.
At each lattice site, a random on-site potential $h_l$ acts on fermions with the strength $W$. Local disorder values
$h_l$ are drawn from the uniform distribution on the interval $\left[-1, 1\right]$. The second term on the right hand  side of Eq.~ (\ref{eq:Hamiltonian}) captures hopping of fermions between the lattice sites. Finally, fermions which occupy neighboring lattice sites, interact with strength $U$.

The open version of the model is described by Eqs.~(\ref{eq:master}- \ref{L}, \ref{eq:Hamiltonian}), with the jump operator operators~\cite{Diehl2008}:
\begin{equation}
	\label{eq:dissipator}
	\begin{gathered}
		L_l=(c_{l}^{\dagger} + c_{l+1}^{\dagger})(c_{l}-c_{l+1}), \quad \forall \gamma_l = \gamma.
	\end{gathered}
\end{equation}
Each one of these operators acts simultaneously on a pair of neighboring sites $(l,l+1)$ and attempts to synchronize the dynamics on these sites by constantly recycling the anti-symmetric (out-of-phase) mode into the symmetric (in-phase) one. 

In the Hamiltonian limit, the MBL regimes are often related to the `integrability'~\cite{Abanin2017} and it was demonstrated that the corresponding Hamiltonians do exhibit level spacing distribution close to the Poissonian pdf; see, e.g., Ref.~\cite{Serbyn2016}.  By gradually increasing disorder strength $W$ from zero, it is possible to tune the Hamiltonian model  (\ref{eq:Hamiltonian} ) from the ergodic phase to the MBL phase  and observe a continuous transformation of pdf $P(s)$ from Poisson to Wigner-Dyson distribution~\cite{Serbyn2016}.

The MBL and ergodic regimes in the open variant of the model, Eqs.~(\ref{eq:Hamiltonian}-\ref{eq:dissipator}), were classified 
in terms of properties of its asymptotic density matrix, $\rho_{a}$, $\mathcal{L}\rho_{a}=0$~\cite{Vakulchyk2018}.
For $J=U=1$ and $\gamma=0.1$ (our choice here too) it was found that the model undergoes a sort of the MBL transition at $W_{\mathrm{MBL}}\backsimeq 3 \div 4$ (the threshold depends also on the model size $M$)~\cite{Vakulchyk2018}. This is in agreement with the MBL threshold $W_{\mathrm{MBL}}\backsimeq 3.5\ldots3.8$ found for the Hamiltonian regime~\cite{Pal2010}.


\section{\label{sec:results} Results}

To calculate LE,  we use initial state  $\varrho(0) = |\psi_0\rangle\langle\psi_0|$, $|\psi_0\rangle = |1010...10\rangle$ for both models.

\begin{figure}[t]
	\begin{center}
		\includegraphics[width=0.85\columnwidth,keepaspectratio,clip]{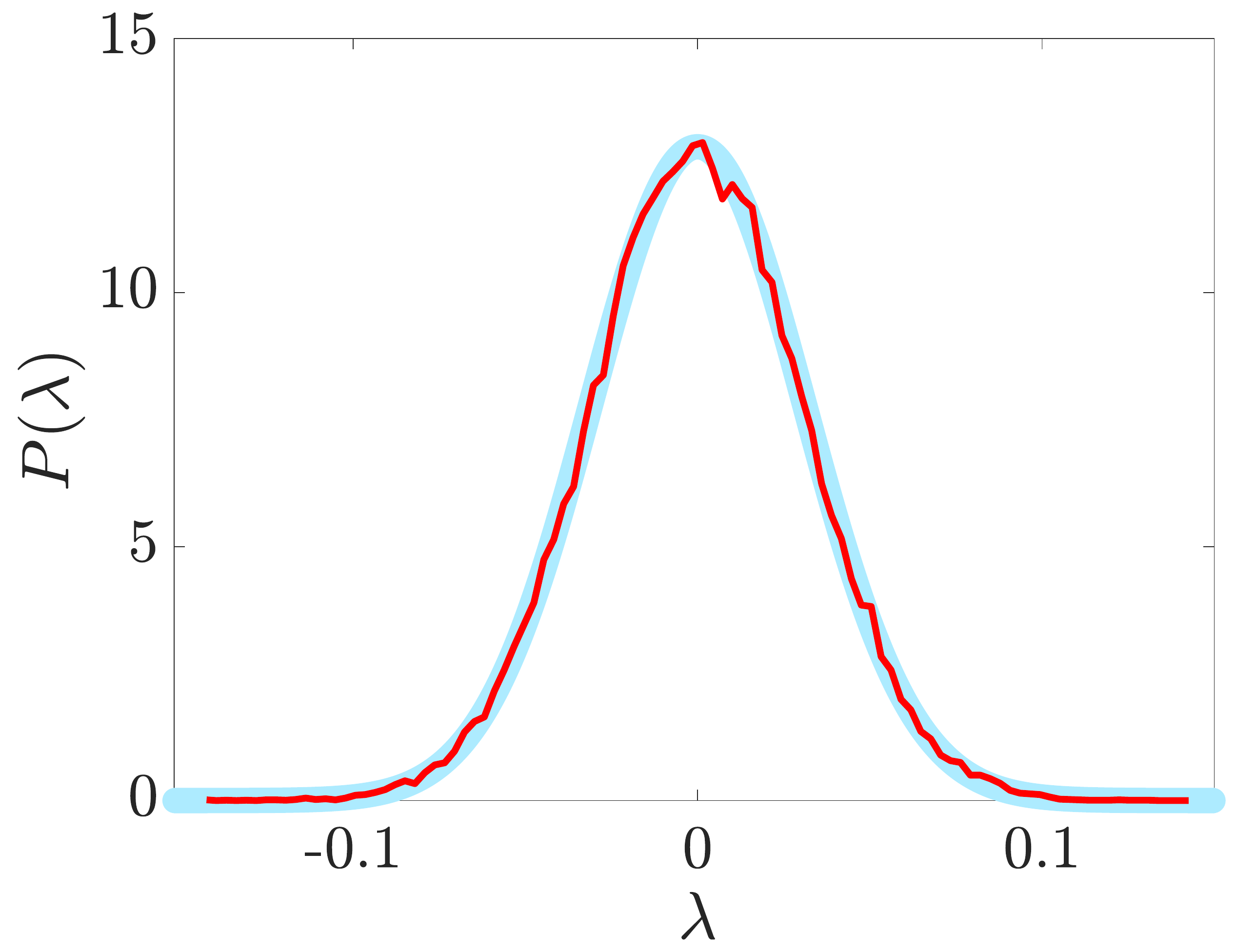}
		\caption{
		    Distribution of largest Lyapunov exponent for the integrable model, Eqs.~(\ref{eq:master}- \ref{L}, \ref{eq:integ}), with $N=7$ spins. Random GOE matrix $O$ was used to calculate LE values. 
		    The initial value of the difference between the trajectories is $\Delta_0 = 10^{-6}$ and the period between consecutive renormalizations  is $\tau=5$.  $10^3$ realizations of $O$ are used  and for each of them $50$ pairs of trajectories were integrated to sample the distribution (thin red line). Normal distribution  $\mathcal N(0,\sigma^2)$ (thick blue line) has zero mean and standard deviation $\sigma = 0.031$.
        }  
		\label{fig:1}
	\end{center}
\end{figure}

We start with the integrable model which constitutes a good test case. The spectrum of its Lindbladian is simple: All jump operators are Hermitian and identical and therefore the spectrum of $\mathcal{L}$ is $M-1$-fold degenerate and purely real. We turn now to the Lyapunov exponents.  

We consider a chain with $M=7$ sites. A randomly drawn GOE matrix $O$
of size $N=2^7$ is used as an observable to sample LEs  according to Eq.~(\ref{eq:le_obs}). We set the initial value of the difference between the trajectories to $\Delta_0 = 10^{-6}$ and the time  of integration before resetting to $\tau=5$. We allow both trajectories to evolve up to time $t_0 = 10^2$, and then follow the dynamics of base, $\psi_{b}(t)$, and perturbed, $\psi_v(t)$, trajectories for time $t = 10^2\tau$. We used $10^3$ random realizations of observable $O$ and for each one of them we run $50$ pairs of trajectories to calculate the Lyapunov exponents ($5 \cdot 10^4$ values in total). The results of the simulations are presented on Fig.~1.

The main observation is that pdf of LEs has a form of a normal distribution with 
zero mean. This implies that the average largest LE is zero and the evolution of the model can therefore be classified as `regular'. The LE pdf 
can be understood if we consider the effective Hamiltonian, $H = -\frac{i}{2}\sum_{l=1}^{M-1}L^{\dagger}L = -\frac{i}{2}\id$. It is an anti-Hermitian operator which induces a monotonous decay of the wave function norm with the constant, state-independent, rate. This means that the corresponding time-continuous evolution does not change the value of $d_k$, Eq.~(\ref{eq:le_growth_factor}), so the latter can only be changed during a jump event. The jump changes $\Delta(t)$ with  equal probabilities to larger and smaller values so that  $\ln\frac{\Delta(t)_{\mathrm{after}} }{\Delta(t)_{\mathrm{before}}}$ is symmetrically distributed around zero. By taking into account that the system perform at least several statistically-independent jumps during  time interval $\tau$, we can state that the Central Limit Theorem applies here and conclude that the quantity defined by Eq.~(\ref{eq:le}) can be  well approximated by a normally distributed random variable.

\begin{figure}[t]
	\begin{center}
		\includegraphics[width=0.95\columnwidth,keepaspectratio,clip]{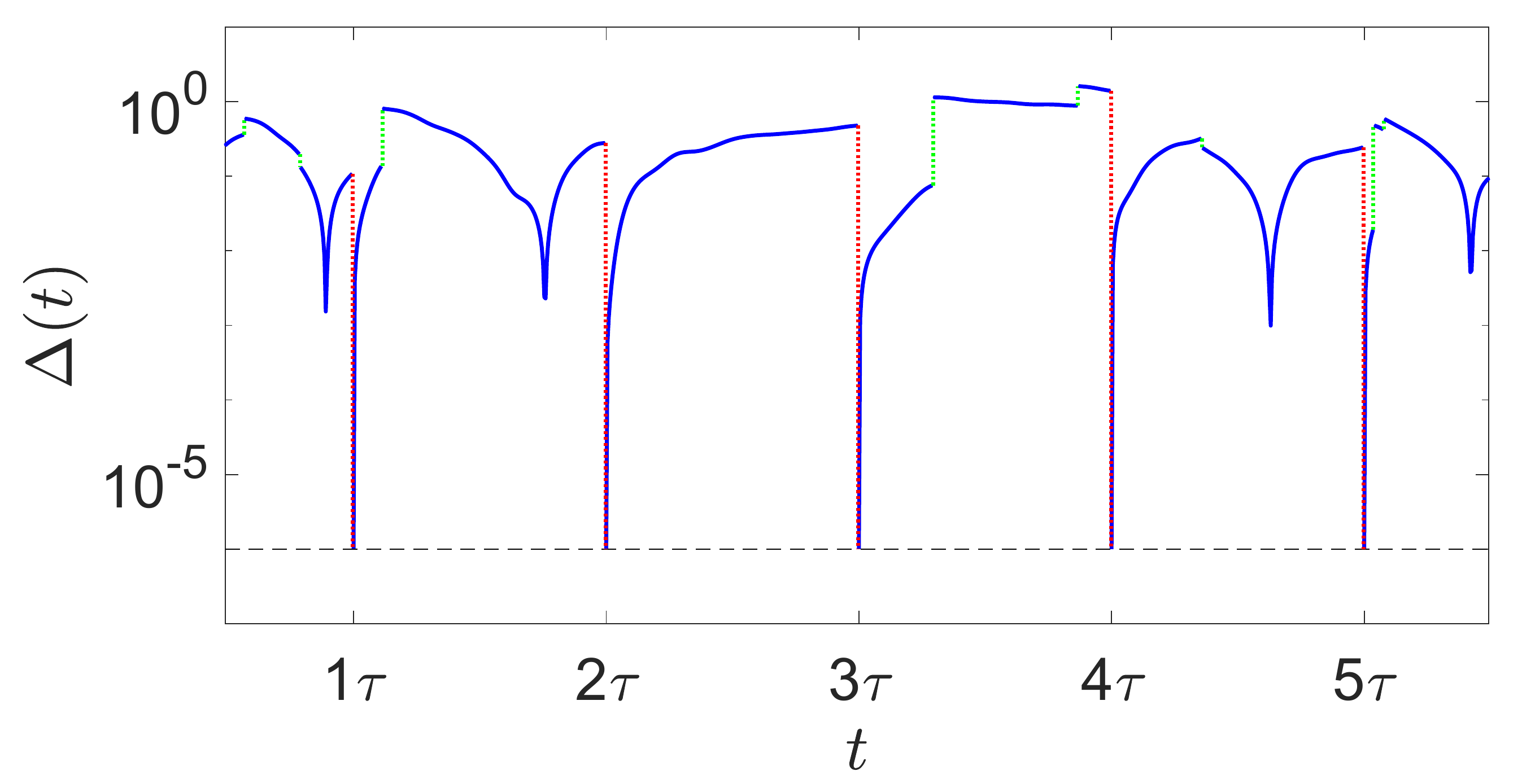}
		\caption{
    		Evolution of the  distance between the base, $\psi_{b}(t)$, and perturbed, $\psi_v(t)$, quantum trajectories, Eq.~(\ref{eq:le_var_obs}) for the MBL model in the ergodic phase. Blue solid lines corresponds to non-Hermitian evolution governed by non-Hermitian Hamiltonian  Eq.~(\ref{eq:qj_hamiltonian}). Green dotted lines corresponds to quantum jumps (note that due to small perturbation values quantum jumps for both trajectories occur almost simultaneously). Red dotted lines mark renormalization events  at times $t_k = k \tau$, $k \in \mathbb{N}$. Other parameters are $\Delta_{0}=10^{-6}$ (indicated with black dashed line), $N=8$, $t=10^3$, $\tau=10$, $W=1$.  
        }  
		\label{fig:2}
	\end{center}
\end{figure}

To calculate LE for the MBL model, Eqs.~(\ref{eq:Hamiltonian} - \ref{eq:dissipator}),  we implemented a high-performance realization of the quantum jumps method \cite{Volokitin2017}. we integrate $10^2$ different trajectories for each disorder realization (we use $10^2$ disorder realizations for each value of $W$). Similar to the previous case,  we allow trajectories to evolve up to time $t_0 = 10^3$ to reach the asymptotic regime, and then we follow the dynamics of both,  $\psi_{b}(t)$ and  $\psi_v(t)$,  trajectories for the time $t = 10^3\tau$.  We again use a small initial distance between trajectories, $\Delta_{0}=10^{-6}$. However, instead of random Hermitian operators, we use the model Hamiltonian as an observable. Finally, we choose resetting time $\tau = 10$. 

Figure~\ref{fig:2} present the evolution of the distance between the base, $\psi_{b}(t)$, and perturbed, $\psi_v(t)$, trajectories of the MBL model  in the ergodic regime,  $W=1$. Values of $d_k$, Eq.~(\ref{eq:le_growth_factor}), are evolving in the interval $10^{-2} - 10^0$ most of the time, which results in a positive  Lyapunov exponent, Eq.~(\ref{eq:le}); see also Fig.~\ref{fig:2}(b).

\begin{figure}[t]
	\begin{center}
		\includegraphics[width=0.95\columnwidth,keepaspectratio,clip]{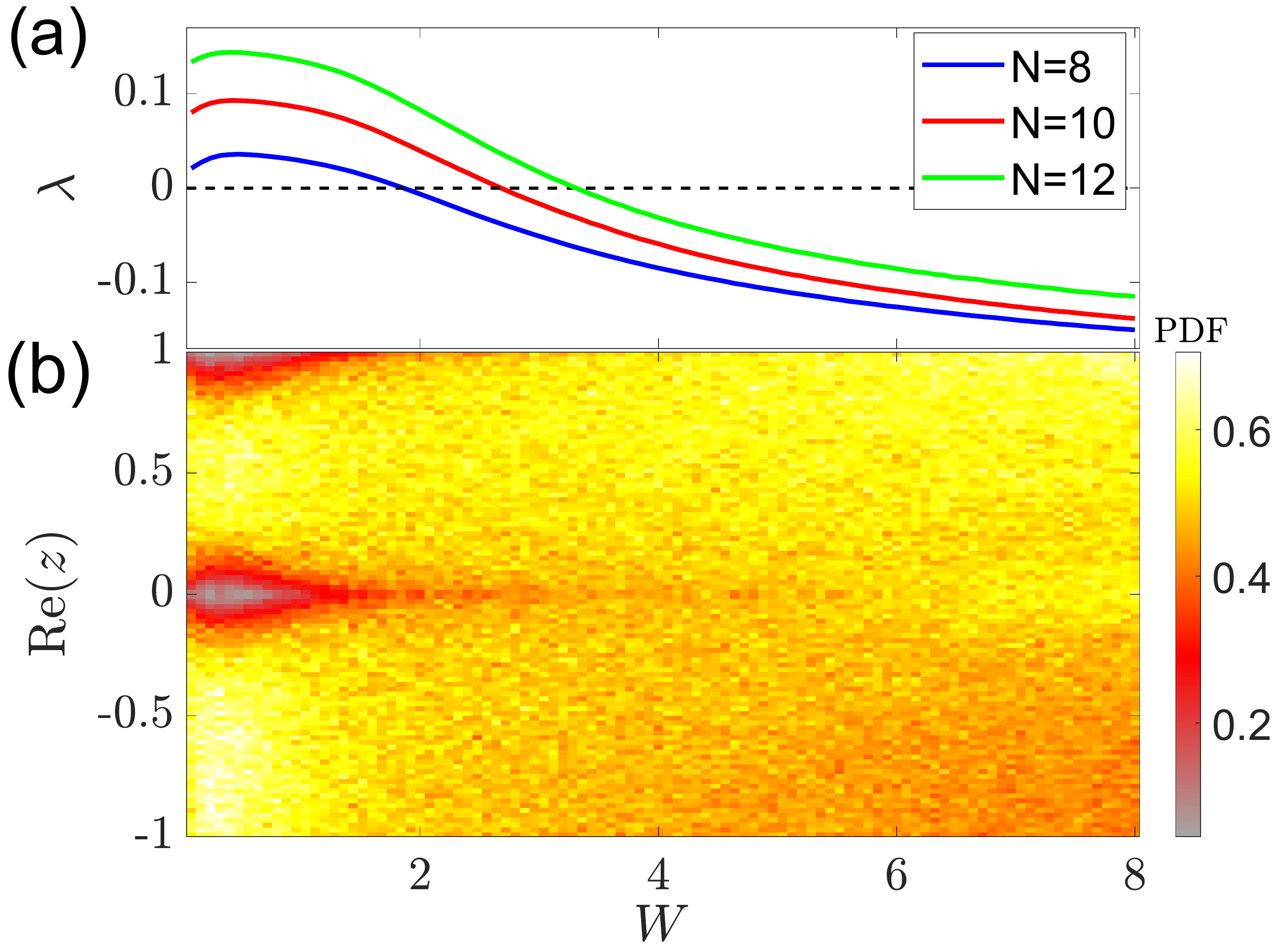}
		\caption{
    		(a) Quantum Lyapunov exponent for the MBL model with $N=8$ (blue), $N=10$ (red), and $N=12$ (green) sites as a function of disorder strength $W$. For each value of $W$ we sample over $10^2$ disorder realizations. For each disorder realization  $10^2$ quantum trajectories are used to calculate the largest Lyapunov exponent. 
    		 (b) One-dimensional probability distribution ($\mathrm{Im}(z)=0$) of complex spacing ratios as functions of $W$. The distributions are obtained by integrating the corresponding two-dimensional distributions in the $\mathrm{Im}(z)$ direction, within the stripe $\vert \mathrm{Im}(z) \vert \leq 0.05$.
        }  
		\label{fig:3}
	\end{center}
\end{figure}

Averaged (over disorder realizations) values of the LEs for different disorder strengths are shown on Fig.~3(a). As expected, LE is positive in the ergodic regime  and becomes negative after the system enters the MBL phase. The transition point depends on the size of the system but it seemingly converges to $W \approx 3.8$, in agreement with the previous findings~\cite{Pal2010,Vakulchyk2018}.


Finally, we sample complex spacing ratio (CSR), Eq.~(\ref{eq:csr}),
for the MBL model. For $W=1$ (ergodic phase), the CSR distribution  exhibits the horse-shoe pattern found for Lindbladians which were defined as chaotic in Ref.~\cite{Sa2020}; see Fig.~4(a). Deep in the MBL regime, $W=20$, the distribution becomes more uniform and approaches a flat disc typical of integrable Lindbladians; see Fig.~4(b). The radial and angular marginal distributions, $\tilde{P}_r(r) = \int d\theta r \tilde{P}(r,\theta) d\theta$ and
$\tilde{P}_{\theta}(\theta) = \int dr \tilde{P}(r,\theta) dr$, where $\tilde{P}(r,\theta)$ is obtained from $P_{\mathrm{CSR}}(z)$ trough $z=re^{i\theta}$, reproduce the results obtained in Ref.~\cite{Sa2020} for chaotic and regular Lindbladians; see Fig.~4(c-d).

To compare the evolution of the PCS distribution  with $W$ and  behavior of the LEs, on Fig.~3(b) we plotted sections of the CSR distribution along the real axis.
It can be seen that the two depletion regions, near $z=0$ and $z=1$, fade with the increase of $W$, and completely vanish beyond the MBL threshold $W_{\mathrm{MBL}} \approx 3.5$.

\begin{figure}[t]
	\begin{center}
		\includegraphics[width=0.95\columnwidth,keepaspectratio,clip]{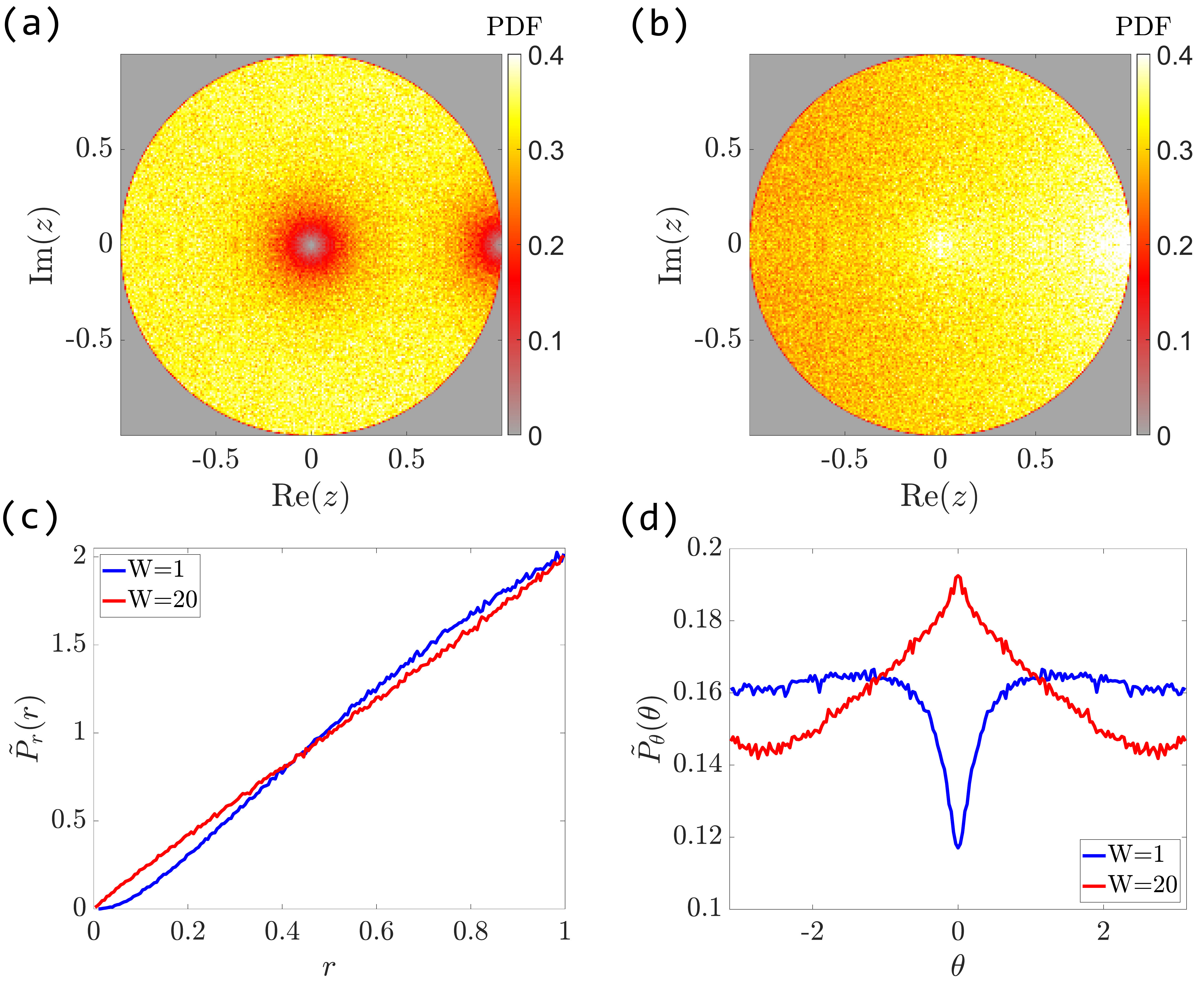}
		\caption{Distribution of complex level-spacing ratios
		    for the MBL model with $N=8$ sites, for $W=1$ (a) and $W=20$ (b).
		    The radial (c) and angular (d)  marginal distributions are obtained based on the two distributions shown on the top panel.
		  }  
		\label{fig:4}
	\end{center}
\end{figure}

\section{\label{sec:conclusions} Conclusions}

By using two open models, we performed a comparative analysis of  two different measures of Quantum Dissipative Chaos (QDC): quantum Lyapunov exponents~\cite{Yusipov2019, Yusipov2020, Yusipov2021} and complex spacing ratio distribution ~\cite{Sa2020}. The first measure is essentially `microscopic' since it is defined by means of quantum trajectories. The evolution of the density matrix, governed by the master equation (1-2), with a generator of the Lindblad form, can be obtained by averaging over an infinite ensemble of quantum trajectories. Complex spacing ratios reflect spectral properties of Lindbladians, generators of open quantum evolution, which act on the density matrix of an open system. This approach therefore is essentially `macroscopic'.

We find the correspondence between the two measures remarkable. However, establishing a formal relation between these two measures remains a challenge. 

Periodically modulated open systems are promising objects of analysis from the  QDC point of view. In Refs.~\cite{Yusipov2019, Yusipov2020, Yusipov2021} the idea of quantum Lyapunov exponents was illustrated by using  this type of models.
However, at the moment it is not cleat how to generalize the idea of different statistics of complex spacing ratios as indicators of DQC to this case as it is not evident what are the corresponding Random Matrix ensembles which should be used as test cases to define `chaotic' and `regular' regimes.

\begin{acknowledgments}
The authors acknowledge support of the Russian Science Foundation through Grant No. 19-72-20086. Numerical simulations were performed on on the supercomputer “Lomonosov-2” of the Moscow State University. 
\end{acknowledgments}

\section*{Data Availability}
Data sharing is not applicable to this article as no new data were created or analyzed in this study.

\nocite{*}

\bibliography{ms}

\begin{thebibliography}{30}
\bibitem [{rem(r  1)}]{remark}%
  \BibitemOpen
  \href@noop {} {} (\bibinfo {year} {In what follows, we scale all relevant
  parameters with respect to the Planck constant to keep $\hbar =
  1$})\BibitemShut {NoStop}%
\end{thebibliography}

\end{document}